\documentclass[twocolumn,english,pra]{revtex4}
\usepackage[T1]{fontenc}
\usepackage[latin9]{inputenc}
\usepackage{amsmath}
\usepackage{graphicx}
\usepackage{amssymb}
\usepackage{esint}

\makeatletter
\@ifundefined{textcolor}{}
{%
 \definecolor{BLACK}{gray}{0}
 \definecolor{WHITE}{gray}{1}
 \definecolor{RED}{rgb}{1,0,0}
 \definecolor{GREEN}{rgb}{0,1,0}
 \definecolor{BLUE}{rgb}{0,0,1}
 \definecolor{CYAN}{cmyk}{1,0,0,0}
 \definecolor{MAGENTA}{cmyk}{0,1,0,0}
 \definecolor{YELLOW}{cmyk}{0,0,1,0}
 }

\makeatother

\makeatother

\usepackage{babel}

\begin{document}

\title{Mean-field study of itinerant ferromagnetism in trapped ultracold
Fermi gases: Beyond the local density approximation}

\author{H. Dong$^{1,2}$, Hui Hu$^{1}$, Xia-Ji Liu$^{1}$ and Peter D. Drummond$^{1}$}

\affiliation{$^{1}$ARC Centre of Excellence for Quantum-Atom Optics, Centre for
Atom Optics and Ultrafast Spectroscopy, Swinburne University of Technology,
Melbourne 3122, Australia }

\affiliation{$^{2}$Institute of Theoretical Physics, The Chinese Academy of Sciences,
Beijing 100190, China}
\begin{abstract}
We theoretically investigate the itinerant ferromagnetic transition
of a spherically trapped ultracold Fermi gas with spin imbalance under
strongly repulsive interatomic interactions. Our study is based on
a self-consistent solution of the Hartree-Fock mean-field equations
beyond the widely used local density approximation. We demonstrate
that, while the local density approximation holds in the paramagnetic
phase, after the ferromagnetic transition it leads to a quantitative
discrepancy in various thermodynamic quantities even with large atom
numbers. We determine the position of the phase transition by monitoring
the shape change of the free energy curve with increasing the polarization
at various interaction strengths. 
\end{abstract}
\maketitle

\section{Introduction}

Recent experimental progress with Feshbach resonances in ultracold
atomic Fermi gases has created opportunities to investigate long-standing
many-body problems in condensed matter physics. One interesting issue
is the problem of itinerant ferromagnetism in two-component (spin-1/2)
Fermi gases with \emph{repulsive} interactions. The study of itinerant
ferromagnetism in condensed matter physics is a fundamental problem
which has an extensive history, dating back to the basic model proposed
by Stoner \cite{StonerPRSLA1938}. However, the phase transition theory
of itinerant ferromagnetism is still qualitative. It is thought that
a Fermi gas with repulsive interactions may simulate the Stoner model
and therefore could undergo a ferromagnetic phase transition to a
spin-polarized state with increased interaction strength \cite{Salasnich2000,SogoPRA2002,MacDonald2005PRL,WuPreprint,LeBlancPRA09}.
This is a result of the competition between the repulsive interaction
and the Pauli exclusion principle. The former tends to induce polarization,
to reduce the interaction energy, while the latter prefers a balanced
system with equal spin populations -- and hence a reduced kinetic
energy. Above a critical interaction strength, the reduced interaction
energy for a polarized Fermi gas will overcome the gain in kinetic
energy. Hence, a ferromagnetic phase transition should occur. Itinerant
ferromagnetism is therefore a purely quantum-mechanical effect which
occurs when the minimum energy is at nonzero magnetization, due to
the Pauli principle.

Recently, an experimental group reported progress in this direction
\cite{ketterle2009Science}, which has attracted intense theoretical
interest \cite{LeBlancPRA09,Conduit2009a,Conduit2009b,Zhai,Fregoso2009,Cui2010,Spilati,Chang}.
By using a \emph{non-adiabatic} field switch to the \emph{upper} branch
of a Feshbach resonance with a positive \textit{s}-wave scattering
length $a_{s}>0$, the experimentalists realized a two-component {}``repulsive''
Fermi gas of $^{6}$Li atoms in a harmonic trap, despite the fact
that the lower branch of the Feshbach resonance is a molecule where
the fermions attract each other. Initial evidence attributed to a
ferromagnetic phase transition has been observed, suggesting that
the transition takes place at $k_{F}^{0}a_{s}\simeq2.2$, where $k_{F}^{0}$
is the Fermi vector of an ideal gas at the trap center. Earlier experiments
measuring interaction energies \cite{Salomon} also provide qualitative
evidence in this direction.

For a trapped Fermi gas, many physical quantities may be used to characterize
the ferromagnetic phase transition. The most direct one is the density
profile of each component. It was suggested from mean-field theory
that with a large scattering length, the two Fermi components in a
trap should become spatially separated \cite{Salasnich2000,SogoPRA2002}.
This would imply that the majority component stays at the trap center
and the minority is repelled to the trap edge. This spatial inhomogeneity
is evidence of a ferromagnetic phase. However, these inhomogeneous
density profiles or spin domains were not observed in the recent experiments,
which is away from thermal equilibrium due to the dynamical use of
the upper branch of Feshbach resonances. For such a non-equilibrium
state, the onset of phase transition is better characterized by a
suppression of inelastic three-body collisions, together with a minimum
in kinetic energy, and a maximum in cloud size. From this evidence,
a phase transition was experimentally identified at a critical scattering
length of $k_{F}^{0}a_{s}\simeq2.2$ and a temperature of $T\simeq0.12T_{F}$.

On the theoretical side, itinerant ferromagnetism is difficult to
treat quantitatively. The phase transition occurs at large interaction
strengths, where fluctuations can be huge. Currently, even the critical
interaction strength for the transition is still under debate. For
a homogeneous gas, mean-field theory predicts a zero-temperature critical
value $k_{F}a_{s}=\pi/2$ \cite{StonerPRSLA1938}, while a second-order
perturbative calculation \cite{MacDonald2005PRL} predicts a much
lower critical coupling strength of $k_{F}a_{s}\sim1.054$. More accurate
quantum Monte Carlo simulations recently reported even lower critical
values $k_{F}a_{s}\simeq0.8$ \cite{Conduit2009b,Spilati,Chang}.
For the experimental situation with harmonic traps, a local density
approximation (LDA) is extensively used in order to average over the
entire trap. At zero temperature, the LDA calculation within the mean-field
theory predicts $k_{F}^{0}a_{s}\simeq1.84$. A finite but low temperature
causes an increase of the critical interaction strength. At the lowest
accessible temperature in the recent experiments ($T\simeq0.12T_{F}$),
we found that the LDA predicts $k_{F}^{0}a_{s}\simeq1.93$.

In this paper, rather than addressing the challenging problem of calculating
the critical interaction strength, we instead examine the validity
of the widely used local density approximation, by solving the Hartree-Fock
mean-field equations self-consistently for a trapped imbalanced Fermi
gas. The LDA is believed to be applicable in general cases. However,
in extreme conditions such as a spin-imbalanced superfluid Fermi gas
in a strongly anisotropic trap, it necessarily breaks down due to
the strongly distorted superfluid-normal interface, which gives a
large surface energy \cite{Silva,Imambekov,Liu2007PRA}. The same
situation could arise in the spatially phase-separated ferromagnetic
phase. In this respect, while the mean-field approximation used in
this work is not exact, our examination of the LDA may provide useful
insight for future applications of the LDA to more accurate homogeneous
equations of state of a repulsive Fermi gas. On the other hand, beyond-LDA
corrections were recently considered by LeBlanc and co-workers through
the phenomenological inclusion of a surface energy term \cite{LeBlancPRA09}.
Our study may be used to determine accurately the phenomenological
parameters as an input.

Our main results may be summarized in the following. First, we find
a notable disagreement between the self-consistent Hartree-Fock solution
and the LDA result in the strongly interacting ferromagnetic phase
where the two species are spatially separated, even with large atom
number up to $10^{5}$. This discrepancy arises from the non-negligible
surface energy, which become increasingly important at large interaction
strength. Secondly, we calculate various thermodynamic quantities
at different temperatures and spin polarizations. The obtained kinetic
energy, interaction energy, and the atomic loss rate all exhibit a
turning point as an indication for the ferromagnetic phase transition.
We also examine the shape change of the free-energy curves as a function
of spin polarization with increasing the interaction strength, from
which we are able to determine accurately a mean-field critical interaction
strength. We find that this does agree with the LDA prediction, presumably
because surface terms play a smaller role at these interaction strengths.

The rest of the paper is organized as follows. In Sec. \ref{sec:Methods},
we briefly outline the model and the Hartree-Fock mean-field approach.
In Sec. \ref{sec:Numerical-Result}, we present a detailed comparison
of our results with that obtained from the mean-field LDA, as well
as a detailed analysis for various thermodynamic quantities. Sec.
\ref{sec:Conclusions} is devoted to the conclusions and further remarks.

\section{Methods\label{sec:Methods}}

We consider an imbalanced Fermi gas with unequal spin populations
in a spherically trap with an \emph{effective} repulsive interaction,
as a minimum model for these experiments. The system may be described
by the Hamiltonian, \begin{equation}
{\cal H=}\sum_{\sigma}\int d{\bf r}\hat{\Psi}_{\sigma}^{\dagger}\left({\bf r}\right)H_{\sigma}^{s}\hat{\Psi}_{\sigma}\left({\bf r}\right)+U\int d{\bf r}\hat{\Psi}_{\uparrow}^{\dagger}\hat{\Psi}_{\downarrow}^{\dagger}\hat{\Psi}_{\downarrow}\hat{\Psi}_{\uparrow},\end{equation}
 where the pseudo-spin $\sigma=\uparrow,\downarrow$ denotes different
hyperfine states and $\hat{\Psi}_{\sigma}\left({\bf r}\right)$ is
the annihilation Fermi field operator for the spin-$\sigma$ state.
The single-particle Hamiltonian $H_{\sigma}^{s}=-\hbar^{2}\nabla^{2}/(2m)+V\left({\bf r}\right)-\mu_{\sigma}$
and $V\left({\bf r}\right)=m\omega^{2}r^{2}/2$ is an \emph{isotropic}
harmonic trapping potential with frequency $\omega$. The effective
repulsive interaction strength is $U=4\pi\hbar^{2}a_{s}/m>0$ in the
lowest Born approximation. The total number of fermions and the number
difference in different hyperfine states are, respectively, $N=N_{\uparrow}+N_{\downarrow}$
and $\delta N=N_{\uparrow}-N_{\downarrow}$. We have introduced different
chemical potentials, $\mu_{\uparrow,\downarrow}=\mu\pm\delta\mu$,
to account for the number difference or population imbalance.

We use the Hartree-Fock mean-field (MFA) approximation to solve the
Hamiltonian, which amounts to decoupling the interaction term $\hat{\Psi}_{\uparrow}^{\dagger}\hat{\Psi}_{\downarrow}^{\dagger}\hat{\Psi}_{\downarrow}\hat{\Psi}_{\uparrow}=n_{\uparrow}\left({\bf r}\right)\hat{\Psi}_{\downarrow}^{\dagger}\hat{\Psi}_{\downarrow}+n_{\downarrow}\left({\bf r}\right)\hat{\Psi}_{\uparrow}^{\dagger}\hat{\Psi}_{\uparrow}-n_{\uparrow}\left({\bf r}\right)n_{\downarrow}\left({\bf r}\right)$.
Here, $n_{\sigma}\left({\bf r}\right)=<\hat{\Psi}_{\sigma}^{\dagger}\hat{\Psi}_{\sigma}>$
is the density. The Heisenberg equation of motion for $\hat{\Psi}_{\sigma}\left({\bf r}\right)$
then reads

\begin{equation}
i\hbar\partial_{t}\hat{\Psi}_{\sigma}=\left[-\frac{\hbar^{2}}{2m}\nabla^{2}+V\left({\bf r}\right)+Un_{\overline{\sigma}}\left({\bf r}\right)-\mu_{\sigma}\right]\hat{\Psi}_{\sigma}.\end{equation}
 We solve the equations of motion by expanding $\hat{\Psi}_{\sigma}\left({\bf r}\right)=\sum_{j}u_{j\sigma}\left({\bf r}\right)\hat{c}_{j\sigma}\exp\left[-iE_{j\sigma}t\right]$,
where the field operator $\hat{c}_{j\sigma}$ annihilates a spin-$\sigma$
fermion in state $j$ with wave function $u_{j\sigma}\left({\bf r}\right)$
and energy $E_{j\sigma}$. This yields the single-particle Schrodinger
equation,

\begin{equation}
\left[-\frac{\hbar^{2}}{2m}\nabla^{2}+V\left({\bf r}\right)+Un_{\overline{\sigma}}\left({\bf r}\right)-\mu_{\sigma}\right]u_{j\sigma}=E_{j\sigma}u_{j\sigma},\label{eq:sphami}\end{equation}
 where $u_{j\sigma}\left({\bf r}\right)$ are normalized as $\int d^{3}{\bf r}|u_{j\sigma}|^{2}=1$.
The density distribution $n_{\sigma}\left({\bf r}\right)$ of spin-$\sigma$
species can be written as

\begin{equation}
n_{\sigma}\left({\bf r}\right)=\sum_{j}|u_{j\sigma}\left({\bf r}\right)|^{2}f\left(E_{j\sigma}\right),\label{eq:density}\end{equation}
 where $f\left(x\right)=1/\left[\exp\left(\beta x\right)+1\right]$
is the Fermi distribution function at an inverse temperature of $\beta=1/(k_{B}T)$.
The chemical potential is determined through the number of atoms for
the two species: \begin{equation}
\int d{\bf r}n_{\sigma}\left({\bf r}\right)=N_{\sigma}.\label{eq:number}\end{equation}

We self-consistently solve the coupled equations (\ref{eq:sphami})-(\ref{eq:number}).
For computational reasons, we introduce a high-energy cut-off $E_{c}$
to truncate the summation in the density equation (\ref{eq:density}).
For high-lying modes with $E>E_{c}$, we then adopt a semi-classical
approximation by assuming that the wavefunction $u_{j\sigma}$ is
locally a plane wave. We refer to refs. \cite{Liu2007PRA} and \cite{Reidl1999PRA}
for details. The density profile can therefore be written in the form,
\begin{equation}
n_{\sigma}\left(r\right)=\sum_{E_{nl}^{\sigma}<E_{c}}\frac{2l+1}{4\pi r^{2}}|u_{nl}^{\sigma}|^{2}f\left(E_{nl}^{\sigma}\right)+\int_{E_{c}}^{\infty}d\epsilon\xi_{\sigma}f\left(\epsilon\right),\end{equation}
 where $\xi_{\sigma}=m\sqrt{\epsilon+\mu_{\sigma}-V\left(r\right)-Un_{\overline{\sigma}}\left(r\right)}/\left(2\pi^{2}\hbar^{2}\right)$
and $u_{nl}^{\sigma}\left(r\right)$ is the radial wavefunction. For
the given numbers of atoms $N$ and $\delta N$, temperature $T$,
and \textit{s}-wave scattering length $a_{s}$, our self-consistent
iterative procedure runs as follows: (i) Start with a Thomas-Fermi
profile for the density distribution of each species or the previous
determined density distribution $n_{\sigma}\left(r\right)$; (ii)
Solve the equation (\ref{eq:sphami}) for all the radial wave functions
$u_{nl}^{\sigma}(r)$ with energy levels below the cutoff $E_{c}$;
(iii) Calculate the new density profiles, adding both contributions
from low-lying modes and high-lying modes; (iv) Update the chemical
potentials according to the number equation (\ref{eq:number}) and
(v )finally check the convergence by comparing the old and new density
profiles. The above steps are repeated until the difference in the
old and new density profiles decreases to an acceptable level. We
note that the value of the high-energy cutoff should be carefully
examined so that the numerical results are independent of the choice
of $E_{c}$.

For the thermodynamic properties, we determine the entropy \textbf{$S$}
straightforwardly by \begin{equation}
S=-k_{B}\sum_{nl\sigma}\left(2l+1\right)f\left(E_{nl}^{\sigma}\right)\ln f\left(E_{nl}^{\sigma}\right),\end{equation}
 where the summation is restricted to the low-lying modes below the
cutoff $E_{c}$, as the contribution from the high-lying modes is
exponentially small. We also calculate the total energy $E$ using

\begin{eqnarray}
E & = & \sum_{\sigma}\left[\sum_{E_{nl}^{\sigma}<E_{c}}E_{nl}^{\sigma}f\left(E_{nl}^{\sigma}\right)+\int d{\bf r}\int_{E_{c}}^{\infty}d\epsilon\epsilon\xi_{\sigma}f\left(\epsilon\right)\right]\nonumber \\
 &  & +U\int dr4\pi r^{2}n_{\uparrow}\left(r\right)n_{\downarrow}\left(r\right)+\sum_{\sigma}\mu_{\sigma}N_{\sigma}.\end{eqnarray}
 The free energy at finite temperatures is obtained by using $F=E-TS$.

Experimentally, the kinetic energy $E_{{\rm ki}n}$ can be obtained
by measuring the radial width of the ultra-cold gas, after free expansion
without a trapping potential or Feshbach magnetic field. In our numerical
work, we calculate the kinetic energy by subtracting from the total
energy the potential energy $E_{{\rm pot}}=\int d{\bf r}V\left({\bf r}\right)n\left({\bf r}\right)$
and the interaction energy $E_{{\rm int}}=U\int d{\bf r}n_{\uparrow}\left({\bf r}\right)n_{\downarrow}\left({\bf r}\right)$,
namely, $E_{{\rm ki}n}=E-E_{{\rm pot}}-E_{{\rm int}}$. Another observable
quantity is the three-body loss rate, which, in the weakly interacting
regime, may be estimated by the expression \cite{Petrov2003PRA},

\begin{equation}
\Gamma=\Gamma_{0}\lambda^{6}\int d\mathbf{r}n_{\uparrow}\left(r\right)n_{\downarrow}\left(r\right)n\left(r\right).\label{eq:LT}\end{equation}
 Here $\lambda=k_{F}^{0}a_{s}$ and $n\left(r\right)=n_{\uparrow}\left(r\right)+n_{\downarrow}\left(r\right)$.
The prefactor $\Gamma_{0}$ may be determined theoretically. While
its detailed value is not of interest in determining the ferromagnetic
phase transition, we note that $\Gamma_{0}$ depends on the scattering
length and Feshbach resonance properties. As this is a low-density
approximation, finite density effects must also be taken into account
in interpreting three-body loss data as a guide to atomic density
and polarization.

\section{Results and Discussions\label{sec:Numerical-Result}}

In the numerical calculations, we use the {}``trap units'' $\hbar=m=\omega=k_{\mathrm{B}}=1$.
The length, energy and temperature are then measured in units of $a_{ho}=\left[\hbar/\left(m\omega\right)\right]^{1/2}$,
$\hbar\omega$ and $\hbar\omega/k_{B}$, respectively. Furthermore,
we will scale our results using characteristic quantities obtained
from a zero-temperature ideal balanced Fermi gas in an isotropic harmonic
trap. These are the Fermi energy $E_{F}=\left(3N\right)^{1/3}\hbar\omega$,
Fermi temperature $T_{F}=E_{F}/k_{B}$, Thomas-Fermi radius $r_{TF}=\left(24N\right)^{1/6}a_{ho}$
and the peak density $n_{TF}=\left(6N\right)^{1/2}/3\pi^{2}a_{ho}^{-3}$.
In most cases, we set the total atom number $N=10^{5}$ and temperature
$T=0.12T_{F}$, comparable to the parameters used in the recent MIT
experiment. To remove complications due to spin degeneracy, we also
set a small polarization $p=(N_{\uparrow}-N_{\downarrow})/N=0.10$.
We use a cut-off energy $E_{c}=85\hbar\omega$ and solve the Hartree-Fock
mean-field equation within the subspace $n_{\mathrm{max}}=80$ and
$l_{\mathrm{max}}=120$, which contains all the energy levels with
energy $E<E_{c}$.

In the following, we first compare the Hartree-Fock mean-field density
profiles with the corresponding LDA predictions to examine the validity
of LDA. We then investigate in detail some thermodynamic properties
and determine the critical interaction strength for emergence of the
ferromagnetic state.

\begin{figure}
\includegraphics[width=0.48\textwidth]{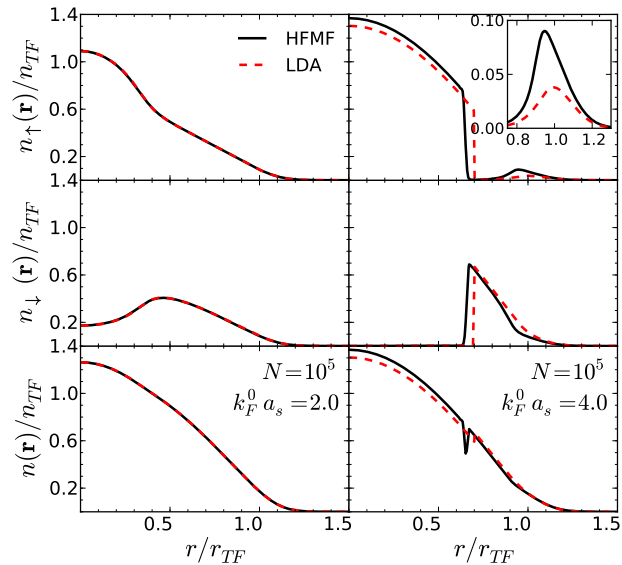}

\caption{(Color online) Spin-up, spin-down and total density profiles of an
imbalanced Fermi gas at $k_{F}^{0}a_{s}=2.0$ (left column) and $k_{F}^{0}a_{s}=4.0$
(right column). The solid and dashed lines refer to the Hartree-Fock
MF and LDA results, respectively. The inset in the upper, right-column
plot highlights the spin-up density profile at the trap edge. The
spin polarization is $p=0.10$ and the number of atoms is $N=10^{5}$. }

\label{fig:densityprofile} 
\end{figure}

\subsection{Hartree-Fock MF versus LDA}

In Fig. \ref{fig:densityprofile}, we present the density profile
of each spin component and the total density profile of the imbalanced
Fermi gas, calculated from Hartree-Fock MF (solid lines) and LDA (dashed
lines) methods with total atom number $N=10^{5}$ and polarization
$p=(N_{\uparrow}-N_{\downarrow})/N=0.10$ at $T=0.12T_{F}$. When
the interaction strength is relatively weak (e.g., before the ferromagnetic
phase transition, $k_{F}^{0}a_{s}=2.0$, left column), the LDA prediction
agrees fairly well with that from the Hartree-Fock MF theory. Similar
excellent agreement is also found in all the physical quantities that
we have considered, including the chemical potential, energy, and
the atom loss rate (see, for example, Fig. \ref{fig:physical Quantities}).
However, when the repulsive interaction is stronger than a critical
value (e.g., $k_{F}^{0}a_{s}=4.0$, right column), there is a significant
discrepancy. In this parameter space, we observe a spatial phase separation
in the density profiles, with majority spin-up atoms occupying the
inner core and minority spin-down atoms repelled to the edge of the
trap. This density separation is revealed clearly by both the LDA
calculations and the Hartree-Fock MF calculations. However, LDA predicts
a sudden change of the density profile of each spin component at $r\simeq0.7r_{TF}$,
due to the neglect of the spatial derivatives of the densities. At
this point, the LDA necessarily fails and one has to resort to the
more reliable self-consistent Hartree-Fock theory. Another notable
discrepancy between LDA and Hartree-Fock MF comes from the trap edge
$r\sim r_{TF}$, where the Hartree-Fock MF theory gives a much larger
spin-up density than the LDA (see, for example, the inset in the right,
upper plot). This is a well-known breakdown of the LDA method due
to small densities occurring at the trap edge. Because of this failure,
as we shall see, the LDA predicts a much smaller atom loss rate than
the Hartree-Fock method. We note also that, for $r\sim r_{TF}$ the
density of both two species becomes very low. There is no spatial
phase separation, as a result of much weaker effective repulsive interactions.

We have also checked a smaller Fermi cloud with $N=10^{3}$ at the
same temperature and polarization. As anticipated, the difference
between Hartree-Fock MF and LDA becomes even more significant.

\begin{figure}
\includegraphics[width=0.48\textwidth]{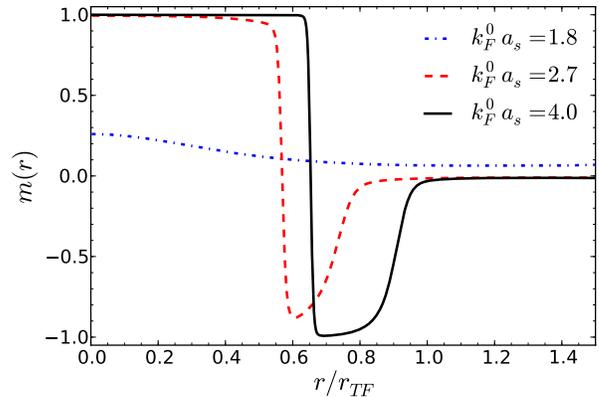}

\caption{(Color online) Magnetization profile or local spin polarization at
different interaction strengths: $k_{F}^{0}a_{s}=1.8$ (dash-dotted
line), $2.7$ (dashed line), and $4.0$ (solid line). These curves
are calculated using the Hartree-Fock MF theory for a total number
of atoms $N=10^{5}$ and a polarization $p=0.10$ at $T=0.12T_{F}$.}

\label{fig:mag} 
\end{figure}

\subsection{Magnetization profile}

As shown above, to reliably characterize the spin density or spin
domains in the ferromagnetic phase, one must include corrections beyond-LDA
by considering surface energy. This was done in the ref. \cite{LeBlancPRA09}
by phenomenologically including a magnetization gradient term.

In Fig. 2, we report the Hartree-Fock MF results for the magnetization
profile, $m\left(r\right)=(n_{\uparrow}\left(r\right)-n_{\uparrow}\left(r\right))/n\left(r\right)$.
Before the transition ($k_{F}^{0}a_{s}=1.8$), the magnetization is
nonzero but small. When the repulsive interaction is larger than the
critical value ($k_{F}^{0}a_{s}=2.7$ and $4.0$), we find a fully
magnetized core, whose size increases with increasing the interaction
strength. As the strength increases, another nearly fully magnetized
domain forms in the vicinity of the trap edge with $m\left(r\right)\approx-1$.
These observations are in qualitative agreement with the theoretical
predictions reported in ref. \cite{LeBlancPRA09}.

\begin{figure}
\includegraphics[width=0.48\textwidth]{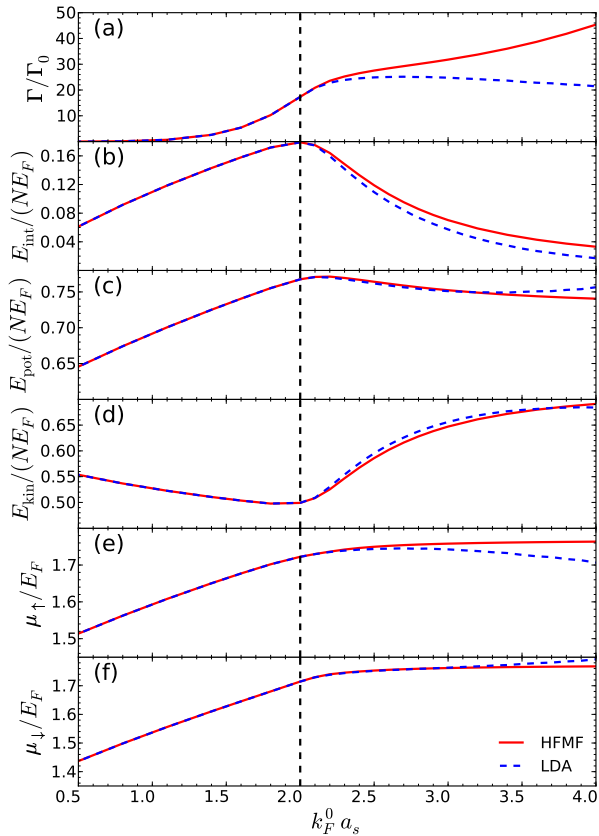}

\caption{(Color online) Atom loss rate (a), interaction energy (b), potential
energy (c), kinetic energy (d), and chemical potentials (e and f)
as a function of interaction strength. The Hartree-Fock MF results
(solid lines) are compared with the LDA predictions (dashed lines).
The vertical dashed line marks the ferromagnetic phase transition.
Here, $N=10^{5}$ , $p=0.10$, and $T=0.12T_{F}$.}

\label{fig:physical Quantities} 
\end{figure}

\subsection{Thermodynamic properties}

We now turn to thermodynamic properties of the imbalanced Fermi gas
with repulsive interactions, and compare our results with the experimental
measurements where available. In Fig. (\ref{fig:physical Quantities}),
we show the atom loss rate (a), interaction energy (b), potential
energy (c), kinetic energy (d) and chemical potentials (e and f).
It is readily seen that LDA and Hartree-Fock MF calculations agree
extremely well with each other for relatively weak interactions below
the critical interaction strength ($k_{F}^{0}a_{s}\sim2.0$), as mentioned
earlier. However, the agreement tends to be worse as the interaction
strength becomes stronger. 

The most significant discrepancy occurs in the atom loss rate of three-body
inelastic collisions. In the Hartree-Fock MF calculation, the loss
rate increases monotonically with increasing interaction strength.
While in the LDA calculations, it increases first but then decreases
when the interaction strength is above the critical value, as illustrated
in Fig. (\ref{fig:physical Quantities}a). This discrepancy lies in
the different prediction of the density tail from the two methods.
It is clear that the loss rate is proportional to the integrated density
profile overlap $n_{\uparrow}\left(r\right)n_{\downarrow}\left(r\right)$$\left(n_{\uparrow}\left(r\right)+n_{\downarrow}\left(r\right)\right)$
over the entire space. For strong repulsion, the spin-up and spin-down
density profiles are essentially not overlapping (i.e., spatially
phase separated), apart from the small regions at the phase-separation
boundary, where the Hartree-Fock MF calculation shows a residue overlap,
and at the edge of the trap, where due to the small density, the effective
repulsion cannot sustain the phase separation further. The atom loss
therefore arises mostly at the phase-separation boundary and at the
trap edge. In these two regions, the LDA fails to give a quantitative
density distribution and hence a reliable prediction for the atom
loss rate. 

However, it should be noted that, the qualitative experimental observation
\cite{ketterle2009Science}, the suppression of the atom loss rate
at large repulsion, agrees with the LDA predictions, but is in contradiction
with our more accurate Hartree-Fock MF calculations. We attribute
the discrepancy to the applicability of equation (\ref{eq:LT}), in
which the strong dependence of the loss rate on the interaction strength
(i.e., scaling as $(k_{F}^{0}a_{s})^{6}$) significantly overestimates
the loss rate above the phase transition. This dependence, however,
is only valid in the weakly interacting regime ($k_{F}a_{s}\ll1$).
A non-perturbative theoretical expression for the loss rate is therefore
needed for large interaction strength, with which we expect that the
Hartree-Fock MF theory would then predict a suppression of the atom
loss rate.

\begin{figure}
\includegraphics[width=0.48\textwidth]{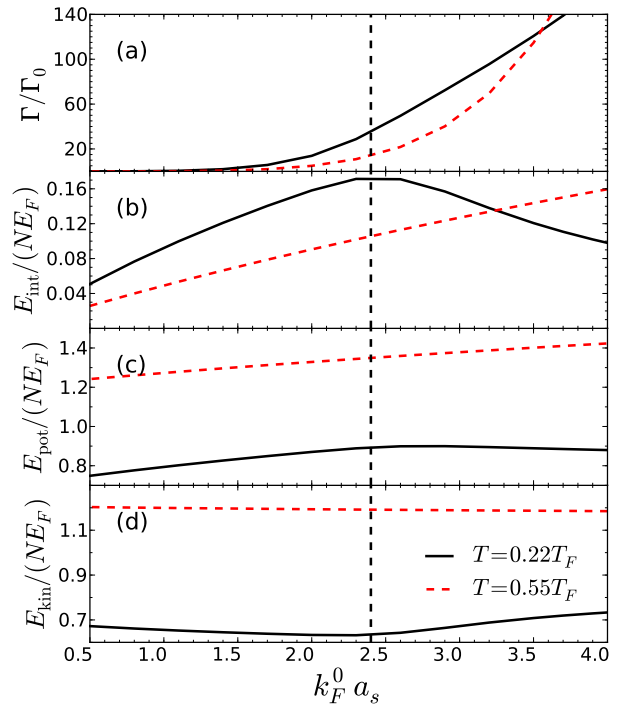}

\caption{(Color online) Atom loss rate (a), interaction energy (b), potential
energy (c), and kinetic energy (d) at different temperatures $T=0.22T_{F}$
(solid lines) and $T=0.55T_{F}$ (dashed line), with a total number
of atoms $N=10^{4}$ and an imbalance ratio $p=0.10$.}

\label{Fig:temperature} 
\end{figure}

There are very clear quantitative differences in the interaction energy,
potential energy and kinetic energy in the ferromagnetic phase predicted
by the two methods (see Figs. (\ref{fig:physical Quantities}b), (\ref{fig:physical Quantities}c),
and (\ref{fig:physical Quantities}d)). These differences are largely
significant once the ferromagnetic domains are formed, and presumably
are due to errors in the LDA treatment of the domain boundaries. Qualitatively,
all the results from the two methods indicate a ferromagnetic phase
transition at about $k_{F}^{0}a_{s}\approx2.0$. The interaction energy
and potential energy show a maximum at the transition, while the kinetic
energy exhibits a minimum. Quantitatively, our mean-field prediction
of the critical interaction strength is $(k_{F}^{0}a_{s})_{c}=1.93$.
This is in good agreement with the MIT experimental finding of $(k_{F}^{0}a_{s})_{c}\simeq2.2$
at the same temperature $T=0.12T_{F}$. 

However, this agreement may simply be a coincidence. Monte-Carlo methods
that take fluctuations into account predict a lower critical strength.
Mean-field theory is not quantitatively reliable at these large interaction
strengths with $k_{F}^{0}a_{s}\geq1$. A possible explanation for
the apparent coincidence is that the experimentally determined critical
strength may not accurately correspond to the true, equilibrium critical
strength. The recent measurements were carried out dynamically. In
this type of non-equilibrium state, critical slowing-down and hysteresis-like
effects will prevent ferromagnetic domains from immediately forming,
and may push the effective critical interaction strength to a higher
value.

The critical interaction strength also depends crucially on temperature.
In Fig. (\ref{Fig:temperature}), we graph the temperature dependence
of the atom loss rate and energies. At higher temperatures, $T=0.22T_{F}$,
the mean-field critical interaction strength increases to $(k_{F}^{0}a_{s})\simeq2.4$,.
This is much \emph{smaller} than the experimental result of $(k_{F}^{0}a_{s})_{c}\simeq4.2$,
at the same temperature. For an even higher temperature, $T\simeq0.55T_{F}$,
there is no non-monotonic behavior in our mean-field results of the
atomic loss and energies and, therefore no indication for the phase
transition anymore. This is in agreement with experimental observation.

We finally consider the chemical potentials as shown in Figs. (\ref{fig:physical Quantities}e)
and (\ref{fig:physical Quantities}f). The chemical potential increases
with interaction strength and nearly saturates in the ferromagnetic
phase. The same saturation was qualitatively observed in the MIT experiment
\cite{ketterle2009Science}.

\begin{figure}[tb]
\includegraphics[width=0.48\textwidth]{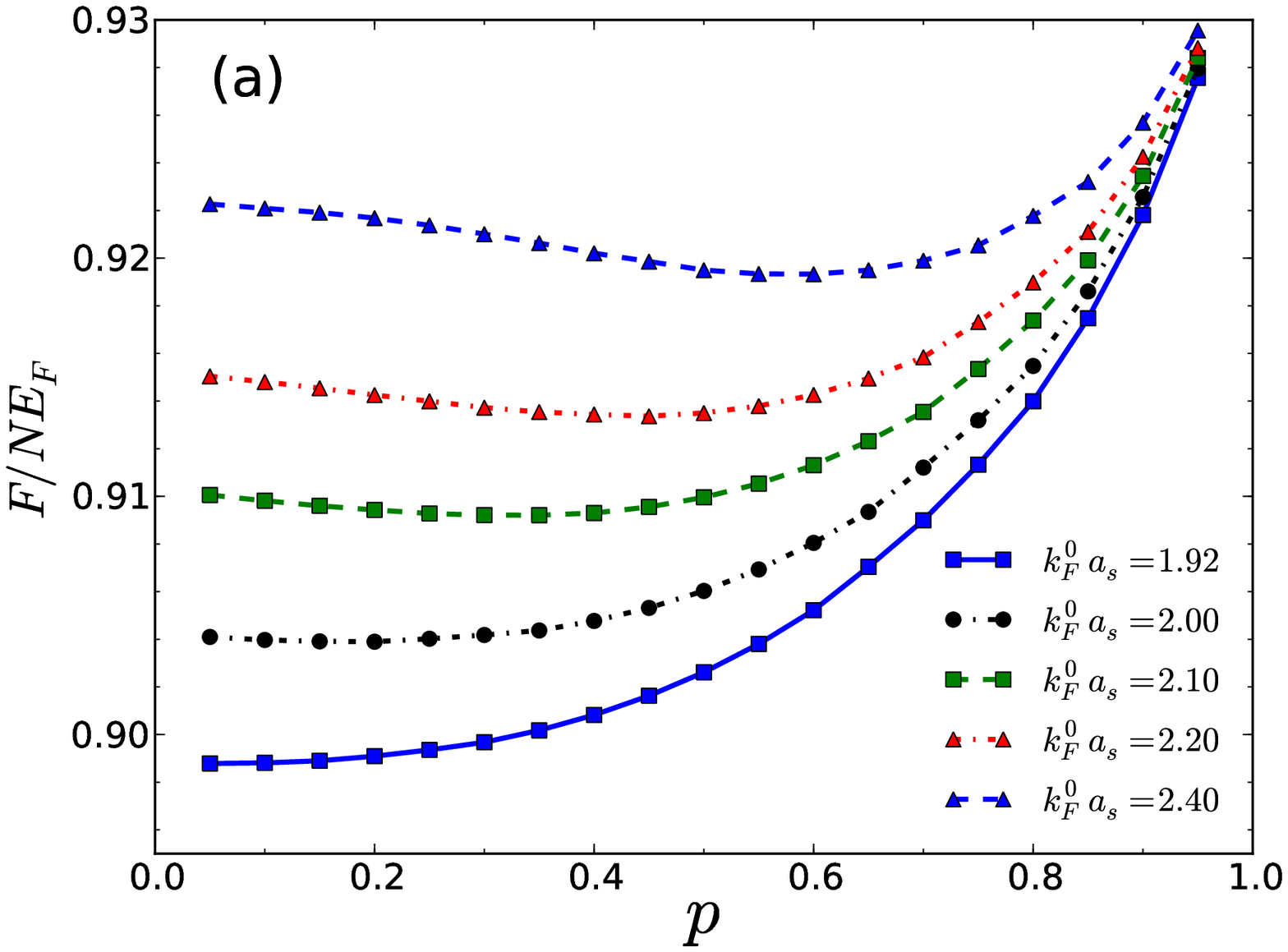}

\includegraphics[clip,width=0.48\textwidth]{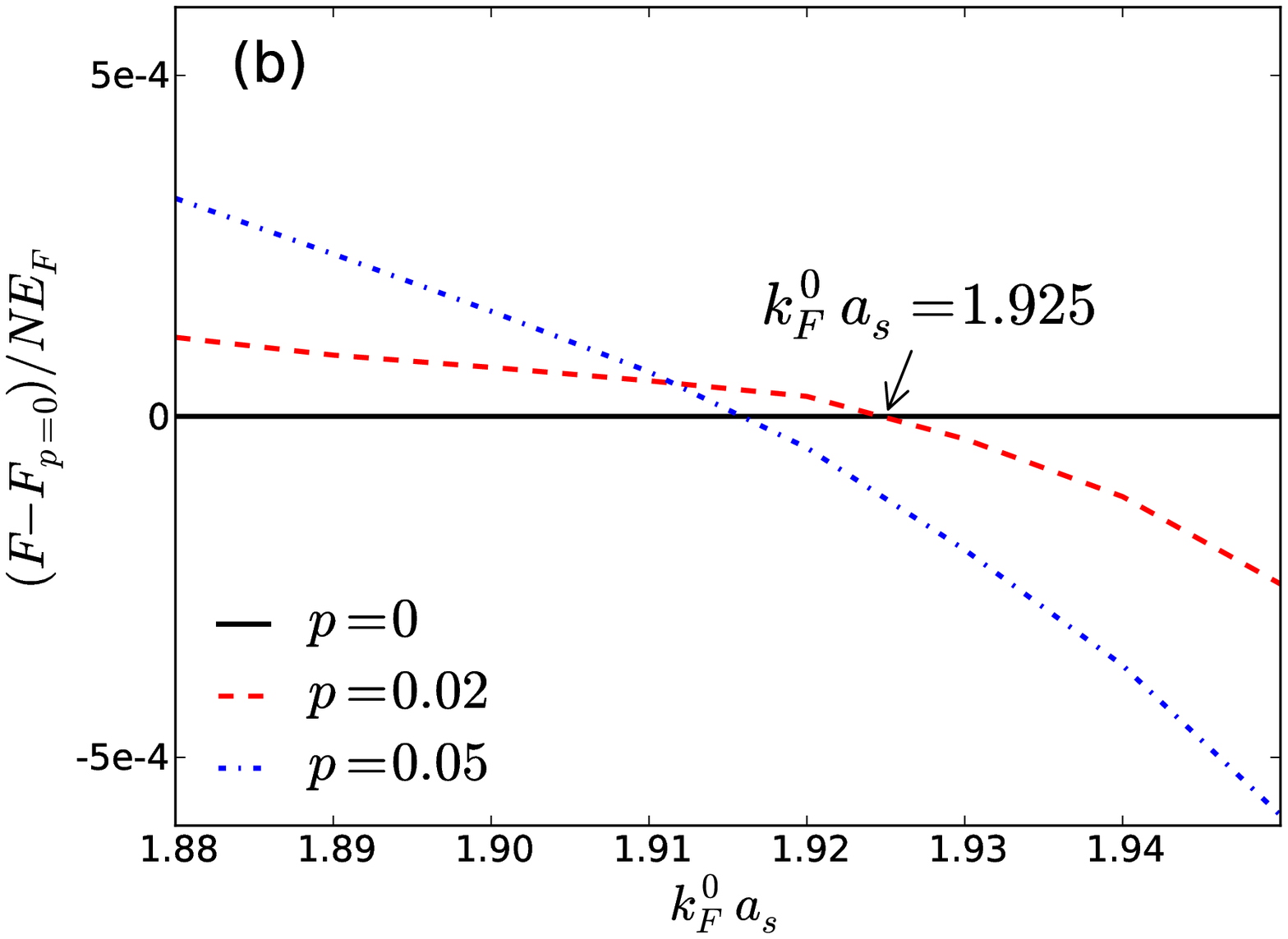}

\caption{(Color online) (a) Free energy as a function of spin polarization
$p$ for different interaction strengths: $k_{F}^{0}a_{s}=1.94,$
$2.00$, $2.10$, $2.20$, $2.40$ and $3.0$. (b) Free energy vs
interaction strength $k_{F}^{0}a_{s}$ for different spin imbalances:
$p=0.05$, $0.10$ and $0.20$. Here, $N=10^{4}$ and $T=0.12T_{F}$.}

\label{fig:enrgy} 
\end{figure}

\subsection{Critical interaction strength from the shape of free energy curve }

At a finite temperature, we can accurately determine the critical
interaction strength from the free energy, which should exhibit a
minimum at non-zero magnetization. We present in Fig. (\ref{fig:enrgy}a)
the free energy as a function of spin polarization at different interaction
strengths. For a weak interaction strength, for example $k_{F}^{0}a=1.92$,
the free energy increases monotonically with spin polarization. Since
the free energy should be symmetric (same) for $\pm p$, we simply
refer this monotonic dispersion to as {}``U'' shape. When the interaction
is large enough, the state with a nonzero spin polarization becomes
energetically favored (see, e.g., $k_{F}^{0}a_{s}=2.4$). Thus, the
dispersion is changed to a {}``W'' shape. As shown in Fig. (a\ref{fig:enrgy}a),
the position of the minimum shifts to large spin polarization with
increasing interaction strength.

To determine the interaction strength accurately for this shape change,
we calculate the free energy $F_{p}$ at three small spin polarizations:
$p=0$, $0.02$ and $0.05$ near $k_{F}^{0}a_{s}=2.0$. In Fig. (\ref{fig:enrgy}b),
and show the free energy with respect to the non-polarization value
$F_{p=0}$. We identify a crossover point at $k_{F}^{0}a_{s}=1.93$,
above (below) which the state with spin polarization is lower (higher)
in free energy. This is the critical interaction strength. For these
atom numbers, our mean-field critical strength is the same as the
LDA prediction for a trapped repulsive Fermi gas at the same temperature.

\section{Conclusions\label{sec:Conclusions}}

In conclusion, we have performed a self-consistent Hartree-Fock mean-field
study of  itinerant ferromagnetism for a harmonically trapped Fermi
gas with strongly repulsive interactions. Our results for the density
profiles, equations of state, and atom loss rate have been used to
examine the validity of the widely used local density approximation.
We have found that the local density approximation works fairly well
below the ferromagnetic phase transition, and gives the same prediction
for the critical coupling strength. However, in the ferromagnetic
or spatially inhomogeneous phase, it gives quantitatively different
predictions for both the equations of state and the atom loss rate,
even for numbers of atoms as large as $N=10^{5}$. This discrepancy
between the local density approximation and the more precise Hartree-Fock
mean-field predictions is apparently due to the breakdown of the local
density approximation at both the phase-separation boundary and at
the edge of the trap.

This failure of the local density approximation is likely to remain
an important issue even if we apply the local density approximation
to the more accurate quantum Monte Carlo data obtained from a \emph{homogeneous}
ferromagnetic Fermi system. For this reason, it would be very useful
to perform an inhomogeneous quantum  simulation in the presence of
a harmonic trap, in order to extend the present work beyond the mean-field
approximation.

Our study may  be useful for calculating the stiffness in the magnetization
gradient term, which is also beyond the LDA. The stiffness has been
computed phenomenologically in ref. \cite{LeBlancPRA09}  using a
Landau-Ginzburg expansion for the excess energy. For the normal-superfluid
interface of a population imbalanced Fermi gas in the unitarity limit,
it was shown that this expansion of gradient terms leads to about
a factor-of-two error at zero temperature, compared with the full
mean-field Bogoliubov-de Gennes calculation \cite{Baur2009PRA}. We
thus anticipate that our Hartree-Fock mean-field calculation will
determine a similar stiffness as derived in  \cite{LeBlancPRA09},
with a similar order of magnitude. However, a complete determination
of the stiffness term is numerically challenging \cite{Baur2009PRA},
especially at finite temperatures.
\begin{acknowledgments}
This work was supported in part by the Australian Research Council
(ARC) Centre of Excellence for Quantum-Atom Optics, ARC Discovery
Project No. DP0984522 and No. DP0984637, NSFC Grant No. NSFC-10774190,
and NFRPC Grant No. 2006CB921404 and No. 2006CB921306.\end{acknowledgments}


\begin{thebibliography}{21}
\bibitem{StonerPRSLA1938} E. C. Stoner, Proc. R. Soc. London. Ser.
A \textbf{165}, 372 (1938).

\bibitem{Salasnich2000}L. Salasnich, B. Pozzi, A. Parola, and L.
Reatto, J. Phys. B: At. Mol. Opt. Phys. \textbf{33}, 3943 (2000).

\bibitem{SogoPRA2002}T. Sogo and H. Yabu, Phys. Rev A \textbf{66},
043611 (2002).

\bibitem{MacDonald2005PRL}R. A. Duine and A. H. MacDonald, Phys.
Rev. Lett. \textbf{95}, 230403 (2005).

\bibitem{WuPreprint}S. Zhang, H.-H. Hung, and C. Wu, arXiv: 0805.3031
(2008).

\bibitem{LeBlancPRA09}J. L. LeBlanc, J. H. Thywissen, A. A. Burkov,
and A. Paramekanti, Phys. Rev. A \textbf{80}, 013607 (2009).

\bibitem{ketterle2009Science}G. B. Jo , Y. R. Lee, J. H. Choi, C.
A. Christensen, T. H. Kim, J. H. Thywissen, D. E. Pritchard, and W.
Ketterle, Science \textbf{325}, 1521 (2009).

\bibitem{Salomon}T. Bourdel, J. Cubizolles, L. Khaykovich, K. M.
F. Magalha, S. J. J. M. F. Kokkelmans, G. V. Shlyapnikov and C. Salomon,
Phys. Rev. Lett. \textbf{91}, 020402 (2003).

\bibitem{Conduit2009a}G. J. Conduit and B. D. Simons, Phys. Rev.
Lett. \textbf{103}, 200403 (2009).

\bibitem{Zhai}H. Zhai, Phys. Rev. A \textbf{80}, 051605(R) (2009).

\bibitem{Conduit2009b}G. J. Conduit, A. G. Green, and B. D. Simons,
Phys. Rev. Lett. \textbf{103}, 207201 (2009).

\bibitem{Fregoso2009}B. M. Fregoso and E. Fradkin, Phys. Rev. Lett.
\textbf{103}, 205301 (2009).

\bibitem{Cui2010}X. Cui and H. Zhai, Phys. Rev. A \textbf{81}, 041602(R)
(2010). 

\bibitem{Spilati}S. Pilati, G. Bertaina, S. Giorgini, and M. Troyer,
arXiv: 1004.1169 (2010).

\bibitem{Chang}S.-Y. Chang, M. Randeria, and N. Trivedi, arXiv: 1004.2680
(2010).

\bibitem{Silva}T. N. De Silva and E. J. Mueller, Phys. Rev. Lett.
\textbf{97}, 070402 (2006).

\bibitem{Imambekov}A. Imambekov, C. J. Bolech, M. Lukin, and E. Demler,
Phys. Rev. A \textbf{74}, 053626 (2006).

\bibitem{Liu2007PRA}X.-J. Liu, H. Hu, and P. D. Drummond, Phys. Rev.
A \textbf{76}, 043605(2007);

\bibitem{Reidl1999PRA} J. Reidl, A. Csordás, R. Graham, and P. Szépfalusy,
Phys. Rev. A \textbf{59}, 3816(1999).

\bibitem{Petrov2003PRA} D. S. Petrov, Phys. Rev. A \textbf{67}, 010703(R)
(2003).

\bibitem{Baur2009PRA}S. K. Baur, S. Basu, T. N. De Silva, and E.
J. Mueller, Phys. Rev. A \textbf{79}, 063628 (2009).
\end{thebibliography}
\end{document}